\documentclass{article}
\usepackage{latexsym}
\usepackage{amssymb}
\usepackage[dvips]{graphics}
\begin{document}
\title{Toward fault-tolerant quantum computation without concatenation}
\author{Eric Dennis}
\maketitle
 
\begin{abstract}

It has been known that quantum error correction via concatenated codes can be done with exponentially small 
failure rate if the error rate for physical qubits is below a certain accuracy threshold.  Other, 
unconcatenated codes with their own attractive features---improved accuracy threshold, local 
operations---have also been studied. By iteratively distilling a certain two-qubit entangled state it is 
shown how to perform an encoded Toffoli gate, important for universal computation, on CSS codes that are 
either unconcatenated or, for a range of very large block sizes, singly concatenated.

\end{abstract}

\section{Codes and computation}

At the end of the long tunnel of experimental quantum computing there is the light of accuracy thresholds 
provided by quantum error correcting codes. The lengthy computations necessary for efficient factorization 
and simulation of quantum systems are all of a sudden possible if error rates for qubits are reduced below 
certain critical values in a sufficiently parallel quantum computer. The quantum codes which give rise to 
this intriguing phase transition work on the same basic principle as their classical precursors---they keep 
information secure by using more physical (qu)bits per logical (qu)bit of the code. What is desired of 
quantum codes is that, as this redundancy is increased, there should be exponential improvement in the 
storage/computation fidelity of the code. Once the basic problem of correcting fully quantum errors was 
solved by the advent of quantum codes, a general recipe was obtained for expanding a given few-qubit code to 
achieve these exponential fidelity gains. This recipe generates a new higher level code by mimicking the old 
code, but with the role of physical qubits played by logical qubits of the old code. Detailed methods of 
error correction and computation have been obtained for versions of this recipe with an arbitrary number of 
iterations or ``concatenations''---even when faults might occur in error correction processes themselves 
\cite{1}--\cite{7}.

However, different frameworks for fault-tolerant error correction, \emph{e.g.} topological quantum codes 
\cite{8}--\cite{11}, might prove superior, and more general methods for universal computation are desirable. 
Here, it is shown how to achieve universal fault-tolerant computation by construction of a Toffoli (C-C-NOT) 
gate on encoded qubits, provided the code possesses a normalizer operator (C-NOT) which factors over the 
qubits in a block.

Most known quantum error correcting codes can be defined by a set of operators, the stabilizer, each of 
which fixes every codeword.  For a number of stabilizer codes capable of simple operations, like a bit flip 
$X_a$ or phase flip $Z_a$ on logical qubits $a,b,\ldots$, it is also known how to perform any ``normalizer'' 
operation---\emph{i.e.} one that can be built from a sequence of operations involving only the C-NOT $\dot 
X_{ab}$, $\pi/2$ phase shift, and Hadamard rotation $R_a$.  Normalizer operations alone, however, are 
insufficient for universal quantum computation; a quantum computer with only normalizer operations can be 
simulated in polynomial time by a classical machine.  A genuine quantum computer is realized either by the 
addition of a non-trivial one or two-qubit gate, like a single qubit rotation by an irrational multiple of 
$\pi$, or of a three-qubit gate like the Toffoli. Here, Shor's procedure \cite{1} for performing a Toffoli 
given the ancilla state
\begin{equation}\label{psi} |\psi_3\rangle
\equiv |000\rangle + |001\rangle + |010\rangle + |100\rangle,
\end{equation}
will be rephrased and then a new method will be given for preparing this state, which does not rely on any 
concatenated structure within the code itself.

\section{Toffoli gate from $|\psi_3\rangle$}

It turns out one can perform a Toffoli gate on three qubits $A\,B\,C$ given normalizer operations and three 
ancilla qubits $a\,b\,c$ prepared in the state $|\psi_3\rangle$.  First consider the following construction, 
which uses one ancilla bit $c$. Letting $c$ start as $|0\rangle$, suppose one could perform a ``majority 
vote'' on $A\,B\,c$ so that, for example, $|010\rangle \rightarrow |000\rangle$ and $|110\rangle \rightarrow 
|111\rangle$. Equivalently, one might majority vote, but only carry out the effect on $c$, leaving $A$ and 
$B$ unchanged, so $|010\rangle \rightarrow |010\rangle$ and $|110\rangle \rightarrow |111\rangle$. Now just 
C-NOT $c$ into $C$ and disentangle $c$ from $A\,B\,C$. The result is exactly a Toffoli on $A\,B\,C$.

To majority vote on $A\,B\,c$, one measures $Z_A Z_B$ and $Z_B Z_c$.  If both measurement results are $+1$, 
$A\,B\,c$ are already unanimous.  Otherwise, the measurement results will reveal which bit is the odd-one-
out.  Unfortunately, these measurements have also revealed information about the initial state $A\,B$, in 
general collapsing it, inconsistent with the desired Toffoli gate, a linear operation. The solution is to 
perform a majority vote not directly on $A\,B\,c$ but on three ancilla qubits, which are first entangled 
with $A\,B$.  Here is where $|\psi_3\rangle$ enters.

Given some arbitrary state of $A\,B\,C$, prepare $a\,b\,c$ as $|\psi_3\rangle$ and perform the following 
operations: (I) C-NOT $A$ into $a$ and $B$ into $b$, and (II) majority vote on $a\,b\,c$ (by measuring $Z_a 
Z_b$ and $Z_b Z_c$ and flipping the odd-bit-out if necessary).  Suppose, for example, the measurement 
results from (II) are $Z_a Z_b = -1$ and $Z_b Z_c = +1$.  All but 8 terms will be collapsed away of the 
total $2^3 \times 4 = 32$ terms in the initial 6-qubit state. These 8 terms, as they undergo (I) and (II), 
are (suppressing bra-ket notation):
\[
\begin{array}{rcccr}
                   &  \mbox{I}   &                     &       \mbox{II}                  \\ 
\smallskip
00C_0 100 & \rightarrow & 00C_0 100  & \rightarrow & 00C_0 000 \\
\smallskip
01C_1 001 & \rightarrow & 01C_1 011  & \rightarrow & 01C_1 111 \\
\smallskip
10C_2 000 & \rightarrow & 10C_2 100  & \rightarrow & 10C_2 000 \\
11C_3 010 & \rightarrow & 11C_3 100  & \rightarrow & 11C_3 000
\end{array}
\]
where $C_i = 0,1$. Note that all of the 8 possible bit values for $A\,B\,C$ are equally represented, so that 
all information in the initial superposition of $A\,B\,C$ is preserved (albeit decoherently).  Now C-NOT $c$ 
into $C$.  From the above table, this will flip $C$ iff $A\,B$ are $01$---not iff $A\,B$ are $11$, as 
desired for the Toffoli.  Applying $\dot X_{BC}$ then gives the desired result.  Finally, $A\,B\,C$ must be 
disentangled from the ancillas $a\,b\,c$ to restore the coherence of the original state.  This is 
accomplished by applying $\dot X_{ab}$ and $\dot X_{ac}$ and then measuring $X_a$.  If the result is $+1$, 
$A\,B\,C$ are disentangled.  If $-1$, a phase error on the $A\,B=01$ term has been introduced; it may be 
corrected by applying $X_A \dot Z_{AB} X_A$, where $\dot Z_{AB} \equiv R_{B} \dot X_{AB} R_{B}$ is the 
controlled-phase (C-PHASE) gate.

Had the measurement results for $Z_a Z_b$ and $Z_b Z_c$ been other than $-1$ and $+1$ respectively, as in 
the above example, it is straightforward to determine what gates must be applied in place of $\dot X_{BC}$ 
and $X_A \dot Z_{AB} X_A$.

The Toffoli now just requires preparation of the three-qubit state $|\psi_3\rangle$. First observe that if 
one can prepare
\[
|\psi_2\rangle \equiv |00\rangle + |01\rangle + |10\rangle,
\]
$|\psi_3\rangle$ may be obtained by preparing four qubits $a\,b\,c\,d$ in the state 
$|\psi_2\rangle|\psi_2\rangle$, measuring $Z_b Z_c$, and performing a few simple normalizer operations.  In 
particular, the measurement result $-1$ gives the state
\[
|0010\rangle + |0100\rangle + |0101\rangle + |1010\rangle,
\]
which can be turned into $|\psi_3\rangle|1\rangle$ by applying the C-NOTs:  $\dot X_{ac}$, $\dot X_{db}$, 
$\dot X_{ad}$, $\dot X_{bd}$, and $\dot X_{cd}$ in that order.

\section{ $|\psi_2\rangle$ from $\rho(\alpha_i)$}

Let us define $\rho(\alpha_1,\alpha_2,\alpha_3)$ as the (unnormalized) mixed state
\[
\left[
\begin{array}{cc}
1 &  \alpha_1  \\
\alpha_2  &  \alpha_3 \\
\end{array}
\right]
\]
in the basis $\{|\psi_2\rangle,|11\rangle\}$, where $|\alpha_3| < 1$. It turns out, in the continuum of 
states $\rho(\alpha_i)$, there is nothing special about $|\psi_2\rangle$, obtained as $\alpha_i \rightarrow 
0$.  Being able to prepare any one state $\rho(\alpha_i)$ with $|\alpha_3| < 1$ is sufficient to prepare 
$|\psi_2\rangle$, hence to prepare $|\psi_3\rangle$ and construct a Toffoli gate.

\begin{figure}[!hp]
\centering
\scalebox{.7}{\includegraphics{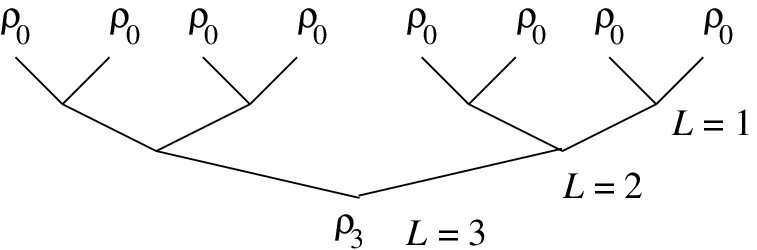}}
\caption{Combining $\rho_0$ states to prepare $\rho_N$ (above, $N=3$).}
\label{tree}
\end{figure}

The state $|\psi_2\rangle$ is prepared by combining two copies of $\rho(\alpha_i)$ through measurement to 
obtain a new mixed state which is closer to $|\psi_2\rangle$ than before, and combining two of these to get 
one still closer, \emph{etc.}, progressively \emph{distilling} $|\psi_2\rangle$ from the initial states. To 
start, prepare qubits $a\,b\,c\,d$ in the state $\rho_0 \otimes \rho_0$, where $\rho_0 = \rho(\alpha_i)$, 
and measure $Z_a Z_c$ and $Z_b Z_d$.  Suppose the results are $+1$ and $+1$.  Now perform $\dot X_{ac}$ and 
$\dot X_{bd}$ to disentangle $c\,d$.  For pure states, this whole process would give 
$|\psi_2\rangle|\psi_2\rangle \rightarrow |\psi_2\rangle|00\rangle$ and $|11\rangle |11\rangle \rightarrow 
|11\rangle|00\rangle$, while either of the initial states $|\psi_2\rangle |11\rangle$ or 
$|11\rangle|\psi_2\rangle$ are inconsistent with the assumed measurement results.  In terms of mixed states, 
this means $\rho_0 \otimes \rho_0 \rightarrow \rho_1 \otimes |00\rangle\langle 00|$ where $\rho_1$ is
\[
\left[
\begin{array}{cc}
1 &  \alpha_1^2  \\
\alpha_2^2  &  \alpha_3^2 \\
\end{array}
\right]
\]
which is exactly $\rho(\alpha_i^2)$. Prepare another $\rho_1$ from two new $\rho_0$ states, and combine the 
two $\rho_1$ states by again measuring $Z_a Z_c$ and $Z_b Z_d$.  Supposing the results are again $+1$ and 
$+1$, $c\,d$ are disentangled, leaving $a\,b$ in the state $\rho_2 = \rho(\alpha_i^4)$. Continuing this 
process through $N$ levels gives $\rho_N = \rho(\alpha_i^{2^N})$.  The whole procedure may be pictured as a 
tree of $\rho_L$ states, joining in pairs from level $L=0$ to $L=N$ (see Fig.\ \ref{tree}).  The 
recursiveness is reminiscent of concatenated codes, but here the complexity appears in the auxiliary 
distillation process, not in the code itself.

The fidelity in preparing $|\psi_2\rangle$ is  
\begin{equation} \label{epsilon}
1-\epsilon \equiv
\frac{\mbox{tr}(\rho_N |\psi_2\rangle \langle \psi_2|)}{\mbox{tr}(\rho_N) \langle\psi_2|\psi_2\rangle} =
\frac{3}{3 + \alpha_3^{2^N}}
\end{equation}
very close to 1 if $|\alpha_3| < 1$.  The number of (logical) qubits used to achieve this fidelity is $\sim 
2^N$, which by (\ref{epsilon}) is $\sim \log \epsilon/\log |\alpha_3|$. This is the same kind of polylog 
scaling desired from the code itself (referring to the scaling of block size with desired failure rate 
$\epsilon$). Finding the number of operations on encoded qubits necessary to prepare $|\psi_2\rangle$ is not 
as easy, since the assumption that all $Z_a Z_b$, $Z_c Z_d$ measurement outcomes are $+1$,$+1$ requires 
repetition of the procedure a number of times before one expects such to occur.

To prepare a single $\rho_L$ state prepare two $\rho_{L-1}$ states and then combine them by measurements. If 
the measurement results are not $+1$,$+1$, just discard these states and keep trying. (This is not an 
optimal procedure, but it will suffice.)  Therefore, if the chances of any one attempt succeeding are 
$P(L)$, the expected number of logical operations $G(L)$ necessary to prepare $\rho_L$ is $\sim 2G(L-
1)/P(L)$. This assumes high confidence in the one pair of measurement results $+1$,$+1$, which should be the 
case since $a\,b\,c\,d$ are \emph{logical} qubits. But even if there is a significant probability 
$\epsilon_\mathrm{m} \gg \epsilon$ for any one measurement result to be in error, the distillation procedure 
can be made robust. Once a $+1$,$+1$ result is obtained, just repeat the measurements a number of times and 
accept the state only if, say, a majority of the results are $+1$,$+1$.  To get $1-\epsilon$ confidence in 
the measurement outcome, one must repeat $\sim \log \epsilon / \log \epsilon_\mathrm{m}$ times.  This 
implies 
\begin{equation} \label{GL}
G(L) \approx \frac{2}{P(L)} \, G(L-1) + \frac{\log \epsilon}{\log \epsilon_\mathrm{m}} \;\;.
\end{equation}

It is not hard to see that $P(L)$ must increase with $L$, since this recursion relation implies that either 
$|\psi_2\rangle$ will quickly begin to dominate successive $\rho_L$ states, in which case $P(L) \rightarrow 
1/3$, or $|11\rangle$ will dominate and $P(L) \rightarrow 1$.  Both of these values are larger than $P(1)$, 
which can be calculated as a function of $|\alpha_3|<1$ but is always bounded from below by 1/4. Iterating 
(\ref{GL}) with this bound gives 
\begin{equation} \label{GN}
G(N) \sim 8^N \frac{\log \epsilon}{\log \epsilon_\mathrm{m}} \sim
\frac{(\log \epsilon)^4}{(\log |\alpha_3|)^3 \log \epsilon_\mathrm{m}} \;.
\end{equation}
Note that $G(N)$ is the total number of logical operations, but these can be done in parallel so that the 
actual distillation time is $\sim N \log\epsilon/\log \epsilon_\mathrm{m} \sim 
\log(|\log\epsilon|)\log\epsilon /\log \epsilon_\mathrm{m}$.  The point is that even with the demand of a 
definite sequence of measurement results, time requirements still scale polylogarithmically with $\epsilon$.  
The crucial fact leading to this scaling is that the probability for getting the measurement results 
$+1$,$+1$ in combining two $\rho_L$ states is finite as $L \rightarrow \infty$.  Thus one can prepare 
$|\psi_2\rangle$, $|\psi_3\rangle$, and execute a Toffoli gate if one can prepare one of the mixed states 
$\rho(\alpha_i)$ with $|\alpha_3| < 1$.

There are multiple ways of obtaining a state $\rho(|\alpha_3|<1)$ for codes which are not too large. In 
fact, Shor's own procedure for preparing $|\psi_3\rangle$ can do so. An alternative method will be presented 
here, applicable to codes possessing a non-trivial normalizer operation (here C-NOT) that is transversal, so 
the encoded operation factors into a number of independent operations on physical qubits. The method works 
by performing a very noisy measurement of the C-NOT operator.

\section{Noisy measurement of C-NOT}

Were it possible to measure C-NOT with high fidelity, one could easily prepare $|\psi_2\rangle = 
\rho(\alpha_i=0)$. It turns out imperfect measurement of C-NOT is still capable of yielding 
$\rho(|\alpha_3|<1)$.

\begin{figure}[!hp]
\centering
\scalebox{.85}{\includegraphics{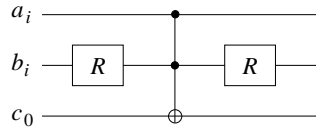}}
\caption{The operation $U$ on physical qubits $a_i\,b_i\,c_0$.}
\label{Ufig}
\end{figure}

For reference, the eigenstates of the C-NOT operator $\dot X_{ab}$ are $|00\rangle$, $|01\rangle$, and
$|10\rangle + |11\rangle$ with eigenvalue $+1$, and $|10\rangle - |11\rangle$ with eigenvalue $-1$.  
Let us first describe a fault-\emph{in}tolerant measurement procedure, that is, one which permits a
single error to spread rampantly throughout a block.  Prepare one physical ancilla bit $c_0$ as
$|0\rangle$ and apply a certain three-bit gate $U_{a_ib_ic_0}$ bitwise over physical bits $a_i$ and
$b_i$ in the blocks encoding $a$ and $b$ (but always using the bit $c_0$). $U$ is shown in Fig.\
\ref{Ufig}.  The first Hadamard rotation causes the Toffoli to flip $c_0$ just if $a_i b_i$ start in the
$-1$ eigenstate $|10\rangle - |11\rangle$ of $\dot X_{a_i b_i}$, and the second Hadamard undoes the
effect on $b_i$.

Apply $U$ bitwise over $a\,b$ and measure $Z_{c_0}$.  The result $Z_{c_0}=\pm 1$ is equivalent to the result 
that $\dot X_{ab} = \prod_{i}\dot X_{a_ib_i} = \pm 1$, so one has effectively measured $\dot X_{ab}$. To 
understand this process in detail, expand the initial state of $a\,b$ in eigenstates of the operators $\dot 
X_{a_ib_i}$ for $i=1,\ldots,n$:
\[
\sum_{\mathbf{x}} C_{\mathbf{x}} |\mathbf{x}\rangle = 
\left( \sum_{w(\mathbf{x})=0} + \sum_{w(\mathbf{x})=1} \right) C_{\mathbf{x}} |\mathbf{x}\rangle
\]
where $|\mathbf{x}\rangle = |x_1\cdots x_n\rangle$ and each $|x_i\rangle$ is one of the four eigenstates 
($x_i = 1,2,3,4$) of $\dot X_{a_i b_i}$.  The right hand sum is rearranged to segregate strings of even and 
odd weight. The weight function $w(\mathbf{x})$ equals the number (mod 2) of ``4''s occurring in the string 
$\mathbf{x}$, $x_i = 4$ corresponding to the $-1$ eigenstate $|10\rangle - |11\rangle$ of $\dot X_{a_i 
b_i}$. Using the transversality of C-NOT and the definition of $w(\mathbf{x})$, one finds $\dot X_{ab} 
|\mathbf{x}\rangle = (-1)^{w(\mathbf{x})} |\mathbf{x}\rangle$.
Thus the sum over strings with $w(\mathbf{x})=0$ is the projection onto the $+1$  eigenspace of $\dot 
X_{ab}$, and the sum with $w(\mathbf{x})=1$ is the projection onto the $-1$ eigenspace. It follows that the 
action of $U=\prod_i U_{a_ib_ic_0}$ is
\[
U |\mathbf{x}\rangle_{ab} |0\rangle_{c_0} = 
|\mathbf{x}\rangle_{ab} |w(\mathbf{x})\rangle_{c_0},
\]
which means measuring $Z_{c_0}$ is equivalent to measuring $\dot X_{ab}$.

This method of measurement is highly sensitive to errors; just one physical bit error can change 
$w(\mathbf{x})$ for an entire string of bits, making the measurement result erroneous.  As the block size 
$n$ gets large, the chances of an  even number of such errors occurring becomes nearly equal to the chances 
of an odd number occurring. Thus the  measurement result tells very little about whether a $+1$ eigenstate 
or a $-1$ eigenstate of $\dot X_{ab}$ has been obtained. This little bit of information, however, turns out 
to be important for preparing $|\psi_2\rangle$. 

As mentioned the above procedure is fault-intolerant, since one physical bit phase error may infect $c_0$ 
and thus spread rampantly throughout the block.  It can be made fault-tolerant by using an ancilla $c$, 
which is not just one bit, but a superposition of $n$ physical bits over all even weight strings (``weight'' 
is now in the sense of counting ``1''s). Such a superposition is prepared as
\[
|\mathrm{even}\rangle_c = \left( \prod_i R_{c_i} \right)(|0\cdots0\rangle_c + |1\cdots1\rangle_c).
\]
The gate $U_{a_ib_ic_i}$ will be applied bitwise across $a\,b\,c$ so that a single error in one block can at 
most spread to one bit in each of the other two blocks.  Acting bitwise on $|x_1\rangle|\;\rangle_{c_1}$ 
through $|x_n\rangle|\;\rangle_{c_n}$, $U$ will flip a number of bits in the initial $c$ state equal (mod 2) 
to exactly $w(\mathbf{x})$.  Thus 
\begin{equation} \label{U}
U |\mathbf{x}\rangle_{ab} |\mathrm{even}\rangle_c =
\left\{
\begin{array}{ll}
|\mathbf{x}\rangle_{ab} |\mathrm{even}\rangle_c & w(\mathbf{x})=0 \\
|\mathbf{x}\rangle_{ab} |\mathrm{odd}\rangle_c & w(\mathbf{x})=1
\end{array}
\right.
\end{equation}
Measuring $Z_{c_i}$ bitwise over $c$ with the result $\prod_{i}Z_{c_i} = \pm 1$ is now equivalent to 
measuring $\dot X_{ab}$ with the result $\dot X_{ab} = \pm 1$.  Note that a single phase error in the 
$n$-bit cat state, or equivalently a bit flip in the sum over even weight strings, will change this sum into 
one over odd weight strings, again altering the measurement result while still projecting the state onto one 
of the eigenspaces of $\dot X_{ab}$. So the measurement procedure is now fault-tolerant, but the measurement 
\emph{result} is still highly sensitive to single bit errors, giving little information about which 
eigenspace the state $|\;\rangle_{ab}$ collapses into.

One can also perform a noisy measurement of the C-PHASE operator $\dot Z_{ab}$.  The action of $\dot Z_{ab}$ 
is just to apply a minus sign if $a\,b$ are in $|11\rangle$, which is unitarily equivalent to $\dot X_{ab}$ 
through the basis change $R_b$.  To measure $\dot Z_{ab}$ first apply $R_b$, then measure $\dot X_{ab}$ by 
the above method, and reapply $R_b$.  These procedures may be adapted, by changing the bitwise operation 
$U$, to noisy measurement of such operators as $\dot X_{ab} \dot X_{cd}$, $\dot Z_{ab} \dot Z_{cd}$, and 
$\dot Z_{ab} \dot  Z_{bc}$.

\section{Preparation of $\rho(\alpha_i)$}

First prepare two logical qubits $a\,b$ as $(|0\rangle + |1\rangle)^2$ and measure $\dot Z_{a b}$ by the 
method given above, making use of an ancilla block $c$. If the measurement result were +1 and all qubits 
were error-free, one would have prepared exactly $|\psi_2\rangle$. But this will be changed by errors 
(\emph{i.e.} decoherence, gate errors, or measurement errors) occurring either to the bits encoding $a$ and 
$b$ or to those of the cat-like ancilla $c$ used in the noisy measurement procedure. In fact $c$ is 
especially vulnerable because it is not protected by any code at all---a single bit error anywhere in $c$ 
can reverse the observed measurement result for $\dot Z_{a b}$.
 
Depending on whether errors are unitary or decoherent, this yields a cohererent or incoherent superposition 
of $|\psi_2\rangle$ and $|11\rangle$, which will be shown to be of the form $\rho(|\alpha_3|<1)$ in either 
the unitary or decoherent case, hence a candidate for distillation.

Phase errors to the bits of $a\,b$ cannot be transmitted to $c$ by the above procedure, so are irrelevant. 
Bit flip errors to $a\,b$ can be transmitted but are equivalent to bit flip errors occurring to the bits of 
$c$ so all errors can be effectively regarded as occurring to $c$ alone. Let us first consider the case of 
(uncorrelated) errors purely decoherent in the Pauli basis $\sigma^m$, so that each qubit $c_i$ suffers no 
error, a phase error, a bit error, or both errors---each with some fixed classical probability.

Phase errors in $c$ can affect only the relative sign of terms in $|\mathrm{even}\rangle_c$ and 
$|\mathrm{odd}\rangle_c$ of (\ref{U}), hence are extinguished once the $Z_{c_i}$ measurements are made. 
Depending on whether $c$ is attacked by an even or odd number of bit errors, the measured eigenvalue of 
$\dot Z_{ab}$ will be inferred either rightly or wrongly from the outcome of the $Z_{c_i}$ measurements. So 
given the result $\prod_{i}Z_{c_i} = +1$, an even number of bits errors will yield $|\psi_2\rangle$ as 
desired; however, an odd number will yield $|11\rangle$ unbeknownst to us. 

If each $c_i$ suffers a bit error with probability $p_i$, the difference between the chances of an even 
number of bit errors and of an odd number is
\begin{equation} \label{prodp}
\left(\prod_{i=1}^n \sum_{x_i = 0,1}\right) (-1)^{x_i} p_i^{x_i} (1-p_i)^{1-x_i}  =  \prod_i (1-2p_i).    
\end{equation}
Given that these two probabilities sum to 1, this implies the preparation procedure will yield not exactly 
$|\psi_2\rangle$, but the state $\rho(0,0,\alpha_3)$ with
\[
\alpha_3= \frac{1-\prod_i (1-2p_i)}{1+\prod_i (1-2p_i)} \approx 1 - 2\prod_i (1-2p_i)
\]
where the last expression holds for large $n$. It thus appears that one cannot tell whether or not 
$|\alpha_3|<1$ for a given ancilla block $c$ if even a few of its bits might have $p_i > 1 /2$. This is true 
even though current codes themselves are completely robust to these ``defective'' bits so long as their 
distribution is suitably uncorrelated and infrequent at the level of the code's threshold error rate. Still 
what has to be considered for the distillation process is not a single ancilla block $c$, giving rise to one 
$|\psi_2\rangle$-like state, but a sequence of such blocks, each with its own set of defective bits and 
consequent value of $\alpha_3 = \alpha_3^{(m)}$, where $m$ runs from $1$ to $2^N$, the number of 
$|\psi_2\rangle$-like states input to the distillation process.

The fidelity in distilling $|\psi_2\rangle$ is that given by (\ref{epsilon}) with $\alpha_3^{2^N}$ replaced 
by
\[
\prod_m \alpha_3^{(m)} \approx
e^{ -2^{N+1} \left\langle \prod_i (1-2p_i) \right\rangle },
\]
where $\langle\cdots\rangle$ is an average over the ensemble of $c$ blocks. Assuming errors are uncorrelated 
between different $c$ blocks, the average factorizes and the distillation fidelity is
\[
1-{\textstyle\frac{1}{3}} \, e^{ -2^{N+1} \prod_i (1-2\langle p_i\rangle) }.
\]
Here only the ensemble averaged bit flip error rates appear. Assuming the locations of defective qubits are 
uncorrelated between different $c$ blocks, their $p_i > 1/ 2$ contributions are simply weighted out in the 
average. This means infrequent defective bits no longer pose a problem, owing to the distributed nature of 
the distillation process. Defining an average error rate $p = (1/n)\sum_i \langle p_i\rangle$, the above 
product is approximately $e^{-2pn}$. In order that the resulting fidelity be comparable to that of the code 
itself, which is $\sim 1 - \exp(-K n^\beta)$ for some power $\beta$ and constant $K$, the number of physical 
qubits used in distillation is roughly
\[
n 2^N \sim n^{1+\beta} e^{2pn},
\]
which puts a limit on the block size $n$ of the code being used, since the number of qubits used in 
distillation should not grow exponentially with block size. Thus $n$ cannot be larger than 
\begin{equation} \label{np}
n \sim \frac{1}{p}\log\frac{1}{p}.
\end{equation}

The opposite case of purely unitary errors is the same in its result. Here, errors comprise a set of unitary 
operators which act on the $c_j$ respectively. Each such operator can be written in the Pauli basis and put 
in the form 
\[
E_j = (A_j\mathbf{1} + iB_j\sigma^z)+ i \sigma^x (C_j\mathbf{1} + iD_j\sigma^z), 
\]
where $A_j,B_j,C_j,D_j$ are real. Assuming low error rates, so $\prod_i |A_i| \gg \prod_i |B_i|$ and 
likewise with $C_i$ and $D_i$ in place of $B_i$, one can show that these errors take $c$ from its prepared 
state $|\mathrm{even}\rangle$ to 
\[
\prod_i E_i|\mathrm{even}\rangle = |\mathrm{even}\rangle + i\tan(\Sigma_C)|\mathrm{odd}\rangle,
\]
where $\Sigma_C \approx \sum_i \tan^{-1}(C_i/A_i)$. This leads to preparation of the state $|\psi_2\rangle + 
i\tan(\Sigma_C)|11\rangle$ in place of $|\psi_2\rangle$. But this state is precisely $\rho(\alpha_i)$ with 
$\alpha_{1,2} = i\tan(\Sigma_C)$ and $\alpha_3 = -\tan^2(\Sigma_C)$, which can be distilled to 
$|\psi_2\rangle$ if $\tan^2(\Sigma_C)<1$. For large $n$, $\Sigma_C$ will be very sensitive to the error 
amplitudes $C_i$, and in practice one would have no way of knowing whether $\tan(\Sigma_C)<1$ or not. This 
is the same problem noted above in the case of pure decoherence, and it too disappears when one realizes 
that the distillation input is not a single state $|\psi_2\rangle + i\tan(\Sigma_C)|11\rangle$ but an 
ensemble of such states each generated by a different set of blocks $a\,b\,c$. The distillation fidelity is 
now given by (\ref{epsilon}) with $\alpha_3^{2^N}$ replaced by 
\[
\prod_{m=1}^{2^N} \tan^2(\Sigma_C) \approx e^{2^{N+1} \langle \log(\tan\Sigma_C) \rangle}.
\]
where $\langle\cdots\rangle$ again averages over the ensemble of $c$ blocks. Using this, the definition of 
$\Sigma_C$, and expanding the logarithm gives $\langle \log(\tan\Sigma_C) \rangle$ as
\begin{equation} \label{ksum}
-\sum_{k=0}^\infty \frac{2}{2k+1} \prod_i \langle  e^{-i2(2k+1)\tan^{-1}(C_i/A_i)} \rangle.
\end{equation}
The factorization follows assuming independence of errors between different qubits in each $c$ block. Now 
expand the exponential in a power series. If the error distributions are all such that the two bit flip 
amplitudes $C_i$ and $-C_i$ are equally likely to occur, the expectation values of the odd terms in the 
power series vanish. (If the distributions are otherwise, one might use the computational basis 
$\{|0\rangle,-|1\rangle\}$ instead of $\{|0\rangle,|1\rangle\}$ for the qubits in half the $c$ blocks, so 
that the same physical error would correspond to the bit flip amplitude $-C_i$ as often as it would to 
$C_i$.) The even terms may then be resummed and the expectation value in (\ref{ksum}) becomes
\[
\left\langle \cos\left( 2(2k+1)\tan^{-1}(C_i/A_i) \right) \right\rangle \equiv 
\cos\left( 2(2k+1)\sqrt{\langle p_i \rangle} \right),
\]
where $\langle p_i \rangle = \langle C_i^2/A_i^2 \rangle$ to lowest order in $C_i/A_i$, hence $\langle p_i 
\rangle$ can be taken roughly as a bit flip error rate. For small $k$ the cosine functions will be close to 
1, and the product over $i$ will be greatest. As $k$ gets large, the cosines will sample their full range 
and the product will be highly suppressed. Therefore the cosine above can be replaced by $\exp(-2(2k+1)^2 
\langle p_i \rangle)$, as if $k$ were always small, giving the main contribution to the sum:
\[
\langle\log(\tan\Sigma_C)\rangle \; \sim \;
-\sum_{k=0}^\infty \frac{2}{2k+1} e^{ -2(2k+1)^2 \sum_i \langle p_i \rangle } 
\; < \; {-2}e^{-2pn},
\]
where $p$ is again the average of $\langle p_i \rangle$ over $i=1,\ldots,n$; this is basically the same 
result as obtained for purely decoherent errors. Thus again (\ref{np}) gives the largest allowed block size 
in the regime where the resources needed for distillation scale polynomially with block size.

\section{Progressive concatenation}

In case higher fidelity is desired of the code than (\ref{np}) allows, the above methods by themselves are 
insufficient and one must resort to concatenation. However, in conjunction with these methods, an 
unconventional, exponentially weaker form of concatenation can be used. A usual concatenated code is self-
similar, the same abstract code (\emph{e.g.} the 7-qubit code) being used at each level in its recursion. 
Here one is free to increase the block size at each level, as long as (\ref{np}) is satisfied level-by-
level. In these ``progressive'' concatenated codes, many fewer levels are necessary given a desired fidelity 
$1-\epsilon$.

In particular, for one error correction algorithm \cite{11} in the context of lattice codes, some reasonable 
parameters are $p = p_c/10 \lesssim 10^{-3}$, so that $n = 1000$ is acceptable by (\ref{np}). Here $\epsilon 
\sim (p/p_c)^{n^\beta}$ where $\beta = \log_9 2 \approx .315$, which gives $\epsilon \sim 10^{-9}$. So, if 
the desired fideltiy is below $1-10^{-9}$, \emph{no concatenation is necessary}. Otherwise, one can begin 
concatenating. 

Consider a single concatenation of a chosen code. Physical qubits with error rate $\epsilon_0=p$ are 
arranged in code blocks of size $n_1$, and these blocks are themselves arranged in blocks of size $n_2$. The 
effective error rate at this higher level is just the failure rate of blocks at the lower level: 
\begin{equation} \label{eps1}
\epsilon_1 \sim (\epsilon_0/p_c)^{Kn_1^\beta},
\end{equation}
where $p_c$, $K$, and $\beta$ come from details of the code being concatenated. The code as a whole has 
failure rate
\begin{equation} \label{eps2}
\epsilon_2 \sim (\epsilon_1^\ast/p_c)^{Kn_2^\beta}.
\end{equation}
where $\epsilon_1^\ast$ includes the effect of storage errors described by $\epsilon_1$ and also gate errors 
associated with the operations necessary to perform error correction. Thus $\epsilon_1^\ast$ will have the 
same form as $\epsilon_1$ in (\ref{eps1}) but with $\epsilon_0=p$ replaced by a physical qubit error rate 
$\epsilon_0^\ast$ including the effect of these additional errors. In other words one must deal not only 
with a storage error threshold but also with a gate error threshold associated with the computations 
necessary for error correction. Of course, if one intends to use the logical qubits stored by the code for 
actual computations, this would be necessary anyway. 

Assuming the effective error rate $\epsilon_0^\ast$ is still below threshold, a single concatenation of the 
code in the above example gives
\[
\epsilon_2 = \left( \frac{(\epsilon_0^\ast/p_c)^{K n_1^\beta} }{ p_c } \right)^{K n_2^\beta} \sim 10^{-830}
\]
where the values $K=1$, $\beta = \log_9 2$, and $\epsilon_0^\ast = p_c/5$ have been used. The block sizes 
$n_1 = 1000$ and $n_2 = 2\cdot10^7$ were chosen to be consistent with $n_L \sim (1/\epsilon_{L-1}^\ast) 
\log(1/\epsilon_{L-1}^\ast)$, \emph{i.e.} the condition (\ref{np}) applied at each level. Thus, as long as 
one does not require a fidelity better than $1-10^{-830}$, a single concatenation is sufficient given the 
above parameters.

Because the block sizes $n_L$ may increase so rapidly, the number $N$ of levels necessary for a desired 
fidelity $1-\epsilon$ scales differently than in usual concatenation, in which $N \sim 
\log(\log(1/\epsilon)) \equiv \log^{(2)}(1/\epsilon)$. For these progressive concatenated codes, $N$ is 
determined self-consistently by $N \sim \log^{(N)}(1/\epsilon)$. One might wonder about the asymptotic 
behavior of thresholds and fidelities as $N \rightarrow \infty$, taking into account the reciprocal effects 
between progressive concatenation and distillation; however, this seems irrelevant given the smallness of 
$\epsilon$ already at $N=2$.

\section{Conclusion and remarks}

It has been shown how to construct a Toffoli gate, the major part of a universal gate set, for any CSS code 
by virtue of its possession of a normalizer operator which factorizes over bits of a code block. For the 
particular code considered above very high fidelity, hence large block size, may require a single 
concatenation, the primary affect of which would be to unify the error rate thresholds for error correction 
and computation. The method presented derives from the ability to perform noisy measurement of the 
(factorable) C-NOT operator on logical qubits. This allows the preparation of a mixed state 
$\rho(|\alpha_3|<1)$, which can then be used in a recursive scheme to distill a certain two-qubit entangled 
state $|\psi_2\rangle$ of logical qubits, which in turn is easily transformed into a three-qubit state 
$|\psi_3\rangle$ that enables the performance of one Toffoli gate on three separate logical qubits.

The point of this construction is to aid the cause of long quantum computations that do not rely on 
concatenated codes. In doing so one trades complexity within the code itself for complexity in processes 
auxiliary to the code with the idea that these auxiliary processes may be performed elsewhere, fostering 
division of labor and economy of scale.

Thanks to Daniel Gottesman, John Preskill, and Atac Imamoglu for comments on this paper.  This work has been 
supported by ARO grant DAAG55-98-1-0366.

\newpage

\end{document}